\documentclass[10pt,twocolumn,floatfix,pra,superscriptaddress]{revtex4-1}
\usepackage[utf8]{inputenc}
\usepackage{amsmath,amssymb,amsthm,mathrsfs,amsfonts,dsfont,amstext} 
\usepackage{graphicx}
\usepackage{physics}
\usepackage{bbold}
\usepackage{bm}
\usepackage{xcolor}

\newcommand{\yso}{Y$_2$SiO$_5$}

\newcommand{\ybi}[0]{$^{171}$Yb$^{3+}$}

\newcommand{\ybiso}[0]{$^{171}$Yb$^{3+}$:Y$_2$SiO$_5$}

\newcommand{\transition}{$^2$F$_{7/2}(0) \leftrightarrow ^2$F$_{5/2}(0)$}

\graphicspath{{./figs/}}

\begin{document}

\raggedbottom	

\title{Coherent optical-microwave interface for manipulation of low-field electronic clock transitions in \ybiso}

\author{L. Nicolas}\email[Email to: ]{louis.nicolas@unige.ch}
\affiliation{Department of Applied Physics, University of Geneva, CH-1211 Geneva 4, Switzerland}

\author{M.~Businger}
\affiliation{Department of Applied Physics, University of Geneva, CH-1211 Geneva 4, Switzerland}
\author{T.~Sanchez~Mejia}
\affiliation{Department of Applied Physics, University of Geneva, CH-1211 Geneva 4, Switzerland}
\author{A.~Tiranov}
\affiliation{Center for Hybrid Quantum Networks (Hy-Q), The Niels Bohr Institute, University of Copenhagen, DK-2100 Copenhagen, Denmark}

\author{T.~Chaneli\`{e}re}
\affiliation{Universit\'e Grenoble Alpes, CNRS, Grenoble INP, Institut Néel, 38000 Grenoble, France}

\author{E. Lafitte-Houssat}
\affiliation{Chimie ParisTech, PSL University, CNRS, Institut de Recherche de Chimie Paris, 75005 Paris, France}
\affiliation{Thales Research and Technology, 91767 Palaiseau, France}

\author{A.~Ferrier}
\affiliation{Chimie ParisTech, PSL University, CNRS, Institut de Recherche de Chimie Paris, 75005 Paris, France}
\affiliation{Facult\'e des Sciences et Ingénierie,  Sorbonne Universit\'e, UFR 933, 75005 Paris, France}

\author{P.~Goldner}
\affiliation{Chimie ParisTech, PSL University, CNRS, Institut de Recherche de Chimie Paris, 75005 Paris, France}

\author{Mikael~Afzelius}
\affiliation{Department of Applied Physics, University of Geneva, CH-1211 Geneva 4, Switzerland}

\begin{abstract}

The coherent interaction of solid-state spins with both optical and microwave fields provides a platform for a range of quantum technologies, such as quantum sensing, microwave-to-optical quantum transduction and optical quantum memories. Rare-earth ions with electronic spins are interesting in this context, but it is challenging to simultaneously and efficiently drive both optical and microwave transitions over a long crystal. In this work, we use a loop-gap microwave resonator to coherently drive optical and microwave clock transitions in \ybiso{}, at close to zero external magnetic field. The low magnetic field regime is particularly interesting for interfacing these spin transitions with superconducting circuits.  We achieve a Rabi frequency of 0.56 MHz at 2.497 GHz, over a 1-cm long crystal. Furthermore, we provide new insights into the spin dephasing mechanism at very low fields, showing that superhyperfine-induced collapse of the Hahn echo signal plays an important role at low fields. Our calculations and measurements reveal that the effective magnetic moment can be manipulated in \ybiso{}, allowing to suppress the superhyperfine interaction at the clock transition. At a doping concentration of 2 ppm and a temperature of $3.4$ K, we achieve the longest spin coherence time of $10.0 \pm 0.4 ~\text{ms}$ reported in \ybiso{}.
\end{abstract}
\maketitle
\section{Introduction}

Solid-state electronic spins coupled to microwave (MW) cavities have been extensively studied for realizing various coherent interfaces for quantum information applications. This is due to the possibility for fast manipulation and their high degree of tunability \cite{morton2018storing}, yet achieving long coherence times with electronic spins remains challenging due to their magnetic sensitivity. Efficient and coherent spin-MW interfaces have been developped for various solid-state electronic spin systems, such as nitrogen-vacancy centers in diamond~\cite{Kubo2010,Zhu2011,Amsuss2011,Ranjan2013,Putz2014}, rare-earth ion doped crystals~\cite{Bushev2011,Probst2013,Chen2016,farina2021coherent,King2021}, phosphorus donors in silicon~\cite{Zollitsch2015,Weichselbaumer2020} and ferri- and ferromagnetic magnons~\cite{Huebl2013,Tabuchi2015,Abdurakhimov2015,everts2020ultrastrong}. Furthermore, electronic spin systems that simultaneously couple coherently to light allow reversible coupling between optical and MW modes, which is a key feature of optical quantum memories \cite{Heshami2016,Rakonjac2021,Ortu2022,Businger2020,Jin2022} and MW-optical quantum transducers ~\cite{Williamson2014,Blum2015,Fernandez2015,Hisatomi2016,bartholomew2020chip}.

In this work, we explore coherent optical and MW coupling in the context of optical quantum memories, based on rare-earth (RE) ion doped crystals. An open challenge on this platform is to realize a broadband ($\geq$~100 MHz) memory while storing the optical photon as a long-lived spin excitation. Broadband operation requires large hyperfine splittings, which can be achieved with electronic spin systems \cite{Businger2020}, i.e. with Kramers RE ions. These, however, are prone to magnetic-induced dephasing due to their large moments. Yet, long electronic spin coherence times can be reached by combinations of moderate to large magnetic fields, sub-Kelvin temperatures and/or ppb-level RE concentrations \cite{Kindem2018o,Rancic2018,Dantec2021}, reaching up to 23 ms in a Er$^{3+}$:CaWO$_4$ crystal at 0.7 ppb concentration at 10 mK \cite{Dantec2021}. Another approach is to exploit zero first-order Zeeman (ZEFOZ) transitions (i.e. clock transitions) that appear at zero magnetic field in Kramers ions having anisotropic hyperfine interactions \cite{Ortu2018,Rakonjac2020,Kindem2020,Chiossi2022}, which can be reached without superconducting coils and with conventional cryo coolers working at 2.5-4 K. Zero-field ZEFOZ transitions have yielded Hahn-echo spin coherence times of up to 2.5 ms on the 655 MHz transition in \ybiso{} \cite{welinski2020coherence} at 10 ppm doping concentration and 3 K. In the MW regime, a Hahn-echo coherence time of 35~$\mu$s was obtained on the 3.369~GHz transition in the optically excited state of \ybi:YVO$_4$ at zero field \cite{bartholomew2020chip}. An important step is to simultaneously demonstrate long spin coherence times in the MW regime $>$1 GHz at the ZEFOZ condition of the ground state, efficient MW manipulation of a spin ensemble, and optical coupling to the same ensemble. In addition, in order to preserve the optical depth for efficient optical quantum memory operation, the coupling must be homogeneous over the length of the crystal.

\begin{figure*}[t!]
	\includegraphics[width=0.95\linewidth]{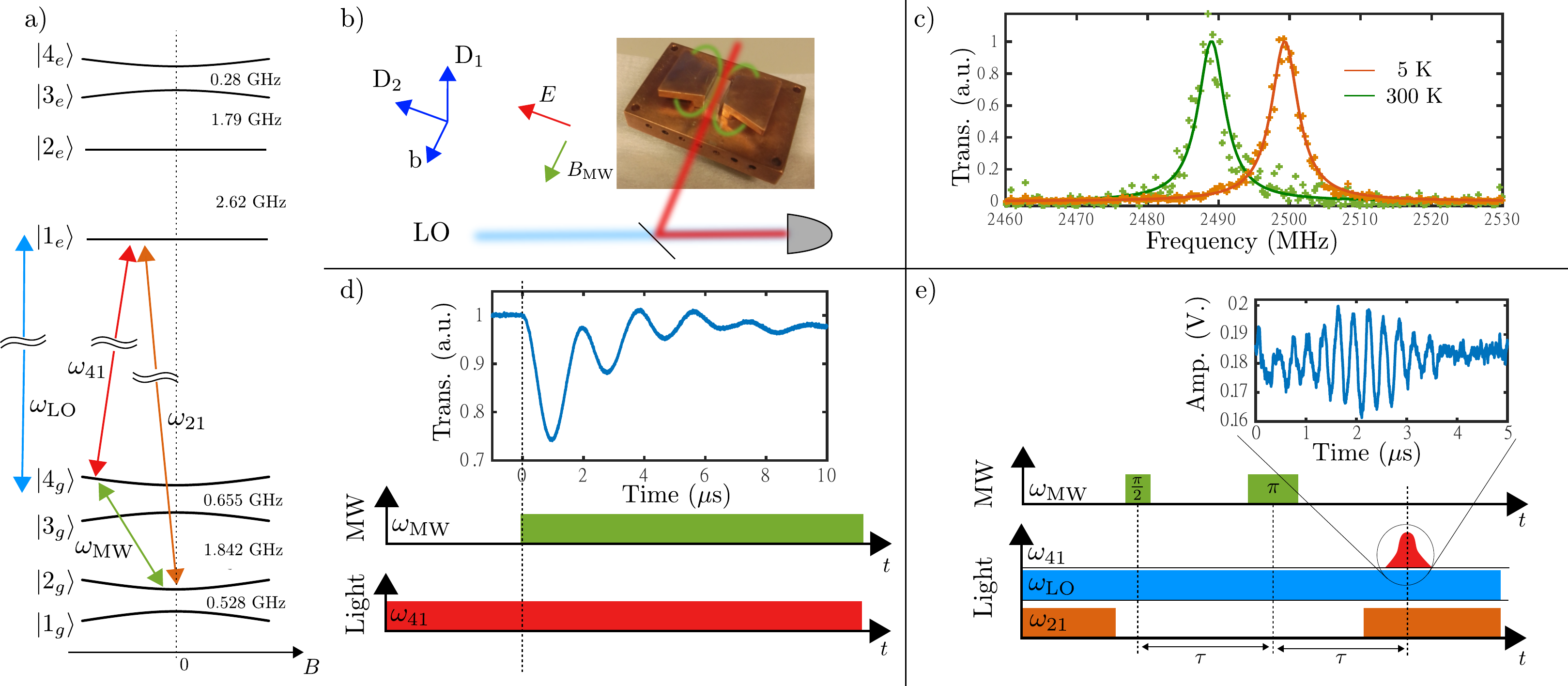}
	\centering
	\caption{a) Energy level diagram for the optical \transition{} transition of site II in \ybiso{}, with relevant optical and MW transition used in the experiments. b) Picture of the loop-gap resonator and the crystal illustrating the common volume excited by the laser $\vec{E}$ and microwave $\vec{B}_{\text{MW}}$ fields, aligned along the D$_2$ and $b$ crystal axes, respectively. c) MW transmission spectra recorded through the resonator at 5~K and 300~K. d) Rabi oscillation time sequence, see text for details, with an example of laser power transmission showing the Rabi oscillations. e) MW Hahn echo time sequence, see text for details. The inset shows an example of the optical RHS echo signal, detected through balanced heterodyne detection at a beat frequency of 3 MHz.}
	\label{fig:resonator}
\end{figure*}

Here, we employ a lumped-element resonator in order to achieve simultaneous optical-MW coupling over a 1-cm long \yso{} crystal. The resonator was tuned to a zero-field ZEFOZ ground-state transition of the \ybi{} ion at 2.497 GHz. The enhanced MW field gives a 0.56~MHz Rabi frequency and Hahn-echo measurements resulted in an electronic spin coherence time of 10 ms, with a \ybi{} concentration of 2 ppm and a temperature of 3.4 K. With small magnetic fields ($<$1~mT), our results provide evidence of strong superhyperfine coupling to neighbouring yttrium ions (mostly $^{89}$Y$^{3+}$ ions which have a 100\% abundance) \cite{Car2018}, suggesting that superhyperfine-induced collapse of the spin echo signal plays an important role in the dephasing, as recently observed with photon echoes in Er-doped \yso{} \cite{Car2020}. This effect can, however, be suppressed at certain angles of the magnetic field and at the ZEFOZ point. Our results shows the possibility of operating an optical quantum memory using any of the MW transitions available in \ybiso{}, over a sample volume compatible with crystals employed for quantum memories.

\section*{Results}

\subsection{Cavity-enhanced spin-microwave interaction}

In this work, we study \ybi{} ions occupying site II of the \yso{} crystal, with the relevant optical transition \transition{} at a wavelength of 978.854 nm \cite{Welinski2016}. The \ybi{} doping concentration is 2 ppm, with an isotopic purity of $\approx 95\%$, and the crystal is cut along the D$_1$, D$_2$ and $b$ axes \cite{Li1992}. The two electronic states have four nondegenerate hyperfine states at zero magnetic field, see Fig. \ref{fig:resonator}(a), due to the  highly anistropic hyperfine interaction in \yso{} \cite{Tiranov2018}. Previous zero-field spin coherence measurements focused on the low-frequency transitions at 528 and 655 MHz transitions \cite{Ortu2018, welinski2020coherence}, which can be rather efficiently excited by a solenoid coiled around the crystal \cite{Ortu2018,Businger2020}. It is interesting, however, to be able to address any MW transition, as other so-called $\Lambda$-systems involving the high-frequency MW transitions could have interesting features for optical quantum memories and transducers. For instance, the electronic spin component of the hyperfine states results in a polarization dependence of the relative transition strengths of the optical-hyperfine transitions (see Supplementary Information), which can be exploited to maximize the coupling strengths of both optical transitions in a $\Lambda$-system. This is in contrast to non-Kramers ions possessing only nuclear hyperfine states, such as Eu$^{3+}$ \cite{Lauritzen2012} and Pr$^{3+}$, where the relative strengths are typically polarization-independent, which severely restricts the number of useful $\Lambda$-systems. Here, we specifically focus on the $\ket{2_g}$-$\ket{4_g}$ transition at $\omega_{\mathrm{MW}}=2496.55$ MHz, because calculations show a particularly low magnetic sensitivity for small bias fields (see Supplementary Information).

To drive the high-frequency MW transition, we designed a loop-gap resonator based on the work by Angerer et al.~\cite{Angerer2016}, which enables efficient and homogeneous driving of the spins over the entire 1-cm length of the \yso{} crystal. In short, the resonator consists of bow-tie type elements, see Fig. \ref{fig:resonator}(b), resulting in a lumped element LC-type electrical circuit where the resonance frequency can be tuned by adjusting the distance between the bow ties and the lid. The design of Ref.~\cite{Angerer2016} was modified by adding optical access through two small holes, allowing optical and MW excitation of a common mode, as shown in Fig. \ref{fig:resonator}(b). The crystal was placed in between the bow ties, with the crystal $b$-axis along the optical beam. The crystal downshifted the resonance frequency by about 60-80~MHz, due to the anisotropic dielectric permittivity of the crystal~\cite{Carvalho2015}. 

In Fig. \ref{fig:resonator}(c), we show transmission spectra acquired with a vector network analyzer (VNA), at a temperature of 300 K and 5 K. Cooling down the resonator typically increased the frequency by $\approx$10 MHz, which was consistent enough between cool downs such that tuning at room temperature was possible. The transmission linewidth is 4.5~MHz, corresponding to a quality factor of $Q = 555$. The resonator was made out of standard copper, without mirror polishing, lowering the $Q$-factor in comparison to Ref. ~\cite{Angerer2016}. However, in this context, the strong coupling regime is not the goal. We use the resonator for a better impedance matching at high MW frequencies and for producing homogeneous oscillating magnetic field $B_{\text{MW}}$ in the crystal volume. For the following measurements, the crystal and resonator were cooled down to about 3.4 K.

The coherent MW driving was measured through optical detection of MW Rabi oscillations. A continuous probe laser tuned to the $\omega_{41}$ transition pumps spins into the $\ket{2_g}$ state, increasing the population difference between the $\ket{4_g}$ and $\ket{2_g}$ states. The transmitted power of the probe laser is recorded on a photodiode, effectively measuring the population in $\ket{4_g}$. The strong $\omega_{41}$ transition allows a high contrast measurement of the population. An amplified MW pulse at 2496.55 MHz was sent to the resonantor, reaching a power of 43~dBm at the MW input of the cryostat. As shown in Fig. \ref{fig:resonator}(d), the MW field induces clear Rabi oscillations between the $\ket{2_g}$ and the $\ket{4_g}$ states, reaching a frequency of $2\pi\times 560$ kHz. Note that the Rabi frequency is comparable to the inhomogeneous spin broadening, which is about 680 kHz. 

The spin coherence time can be measured using a Hahn echo sequence on the MW transition, which can be optically detected using Raman heterodyne scattering (RHS) \cite{Mlynek1983,Ortu2018,everts2020ultrastrong}, see Fig. \ref{fig:resonator}(e). A pulsed probe laser tuned to the $\omega_{21}$ transition first polarizes spins into the $\ket{4_g}$ state. Then the MW Hahn echo sequence is applied, consisting of a 0.42-$\mu$s long $\pi/2$-pulse and a 0.84-$\mu$s long $\pi$-pulse, separated by a time $\tau$. The MW echo is detected by applying another probe pulse at the moment of the echo, which produces a RHS signal on the strong $\omega_{41}$ transition. For an increased sensitivity, the RHS signal field amplitude is detected through a balanced heterodyne detection with a local oscillator (LO) detuned by 3~MHz from the RHS signal, see Fig. \ref{fig:resonator}(e). The spin echo results will be discussed in the following section.

\subsection{Low-field spin properties of \ybiso{}}

The $C_1$ point symmetry of the doping sites in \yso{} results in an effective electronic spin $S=1/2$ of the Yb$^{3+}$ ion, and the isotope \ybi{} has a nuclear spin $I = 1/2$. The effective spin Hamiltonian can then be decomposed into a hyperfine and an electronic Zeeman part \cite{Tiranov2018} (neglecting the weak nuclear Zeeman effect), for the ground ($g$) and electronic level ($e$), respectively,

\begin{equation}
	\label{eq:Heff}
	\boldsymbol{\mathcal{H}} = \mathbf{S} \cdot \mathbf{A}_{g,e} \cdot \mathbf{I} +  \mathbf{B} \cdot  \boldsymbol{\mu}_{g,e}.
\end{equation}

\noindent The electronic spin $\mathbf{S}$ and the nuclear spin $\mathbf{I}$ components are coupled through the hyperfine tensor $\mathbf{A}_{g,e}$. The magnetic dipole moment is $\boldsymbol{\mu}_{g,e} = \mu_\text{B} \mathbf{g}_{g,e} \cdot \mathbf{S}$, where $\mathbf{g}_{g,e}$ is the Zeeman tensor and $\mu_\text{B}$ the Bohr magneton. In a $C_1$ point symmetry, $A_x \neq A_y \neq A_z$, which completely lifts degeneracy at $B=0$ and results in four hyperfine eigenstates $\ket{k_i}$ ($k=1$ to $4$) in each electronic level ($i = g,e$), see Fig. \ref{fig:resonator}(a). At $B=0$, the hyperfine states are completely hybridized in their electronic and nuclear components, such that $\expval{\mathbf{S}} =\expval{\mathbf{I}} = 0$. As a consequence, the expectation value of the dipole moment is zero $\expval{\boldsymbol{\mu}_{g,e}} = 0$, and to first order there is no linear Zeeman effect in either electronic level. The zero first-order Zeeman effect (ZEFOZ) effect at $B=0$ leads to a strong increase in both optical and spin coherence times \cite{Ortu2018}. The zero effective magnetic dipole moment also quenches superhyperfine interactions to first order (see Supplementary Information), thereby quenching the electron spin echo envelope modulation (ESEEM) 
\cite{rowan1965electron,car2018selective,lovric2011hyperfine},  as first observed experimentally in Ref. \cite{Ortu2018}.

\begin{figure*}[t!]
	\includegraphics[width=\linewidth]{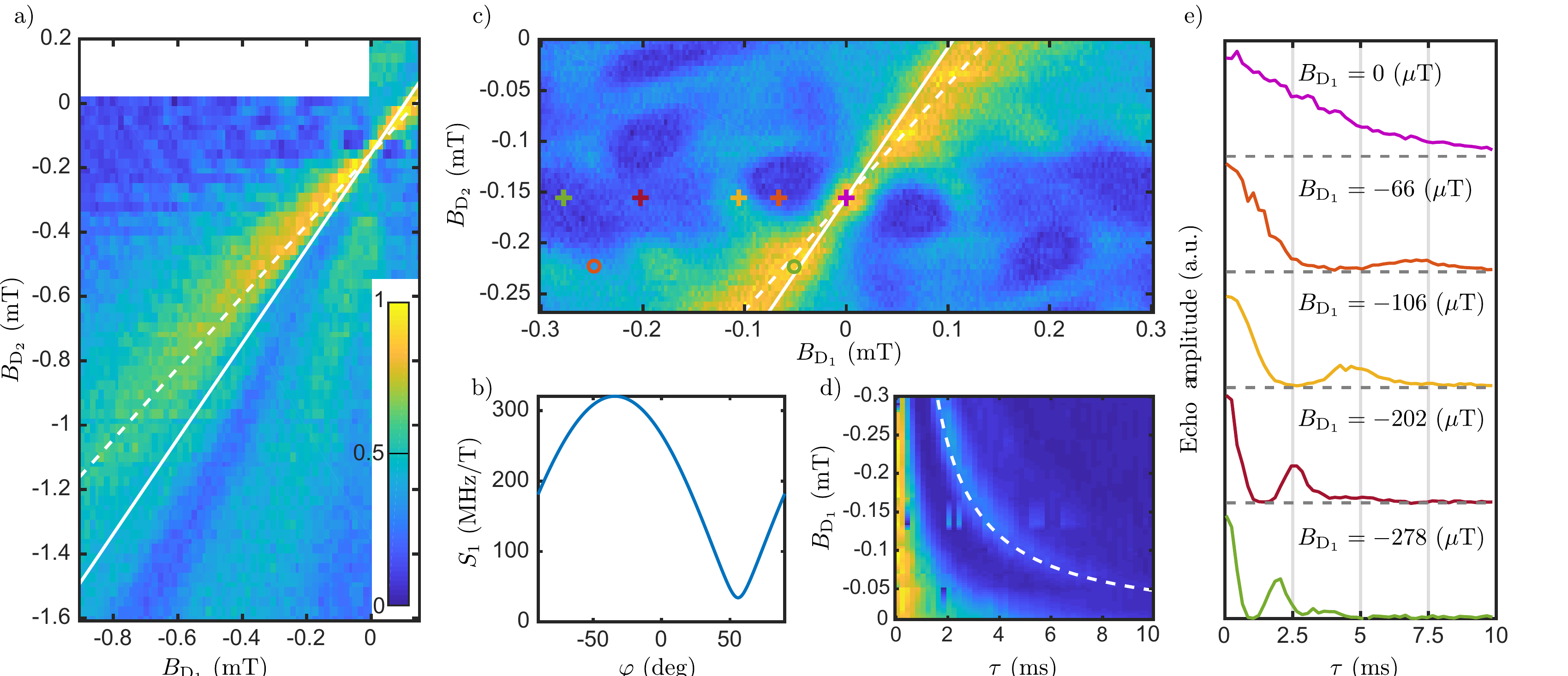}
	\caption{a) Map of the relative Hahn echo amplitude normalized to the maximum amplitude for a range of magnetic fields in the D$_1$-D$_2$ plane, with the pulse separation fixed at $\tau=5$ ms. b) Numerically computed $S_1$ gradient for the 2.497 GHz transition in the D$_1$-D$_2$ plane, as a function of the angle $\varphi$ to the D$_1$ axis. The solid lines in a) and c) show the expected minimum at $\varphi = 55.9^{\circ}$, while the dashed lines show the estimated angle of maximum echo amplitude, at an angle of $\varphi = 48^{\circ}$. They agree within the expected error of the field direction and of the cuts of the crystal surfaces. c) Similar to the plot in a), but showing data taken at a higher resolution in the field step. The color-coded crosses and circles show the field values at which the echo decays shown in e) and Fig. \ref{fig:scanD1_ESEEM}(b)-(c) were recorded, respectively. d) Map of the echo decays as a function of time delay $\tau$, for a range of fields along $B_{\text{D}_1}$ and for fixed $B_{\text{D}_2} = -155~\mu$T. The dashed line shows the yttrium Larmor period. Examples of decay curves from the map are shown in e), from top to bottom $B_{\text{D}_1} = 0, -66, -106, -202, -278 ~\mu$T, indicated with crosses in panel c).}
	\label{fig:scanD1D2}
\end{figure*}

When a weak magnetic field is applied, the states $\ket{k}$ are mixed trough the Zeeman interaction (note that in the following we have omitted the index $i=g,e$ indicating the electronic level). The weakly perturbed states $\ket{l}$ result in a non-zero expectation value $\bra{l} \mathbf{S} \ket{l} \neq 0$, such that each state acquires a weak magnetic dipole moment $ \bra{l} \boldsymbol{\mu} \ket{l} \neq 0$. If the $\mathbf{A}$ and $\mathbf{g}$ tensors are aligned, a good approximation here \cite{Tiranov2018}, then a first-order perturbation calculation shows (see Supplementary Information) that the energy $E_l$ of the perturbed state $\ket{l}$ has a quadratic Zeeman effect, with an associated field-gradient $dE_l/dB_m = \mu_\text{B} g_m \bra{l} S_m \ket{l} $, where $m=x,y,z$ is the direction of the weak magnetic field. It can further be shown that the effective dipole moment component along $m$, for state $\ket{l}$, is given by  $\bra{l} S_m \ket{l} = (1/2) \mu_\text{B} B_m g_m / (E_l - E_{l'})$, where $\ket{l'}$ is the only other hyperfine state such that $\bra{l'} S_m \ket{l} \neq 0$. Note that the zero-field energies $E_l$ can easily be calculated from the $\mathbf{A}$ tensor elements \cite{Tiranov2018}. The direction-averaged, first-order, transition frequency gradient $S_1$ is calculated by taking the difference of the energy gradients of the connected states.

The calculations emphasize the direct connection between the first-order dipole moment and the energy gradient when a weak field is applied, as expected for a classical dipole. Both are zero to first order in the ZEFOZ point, but can also be strongly quenched when applying a weak field, by tuning the field direction to minimize the nominator term $g_m^2$ and to maximize the denominator term $E_l - E_{l'}$, as shown experimentally in Ref. \cite{Ortu2018}. In the following, we will see how both effects change the observed spin dephasing time, and the strength of superhyperfine interaction and the ESEEM effect.

\subsection{Spin echo measurements}

The Hahn echo signal amplitude was measured as a function of magnetic field in the D$_1$-D$_2$ plane, i.e. with $B_{\text{D}_1}$ and $B_{\text{D}_2}$ components, for a fixed pulse separation of $\tau=5$~ms, see Fig. \ref{fig:scanD1D2}(a). The echo signal is strongest in the vicinity of zero applied field, as expected for the ZEFOZ point, but also remains strong along a particular line. The intense line corresponds well to the minimum $S_1$ gradient in the D$_1$-D$_2$ plane, expected at an angle of $\varphi = 55.9^{\circ}$ to the D$_1$ axis (see Fig. \ref{fig:scanD1D2}(b) and Supplementary Information). In addition, one can observe a distinct ESEEM oscillation pattern, which is a sign of dipole-dipole coupling to neighbouring $^{89}$Y$^{3+}$ ions (i.e. superhyperfine interaction).

\begin{figure*}[t!]
	\includegraphics[width=0.95\textwidth, height=3.5cm]{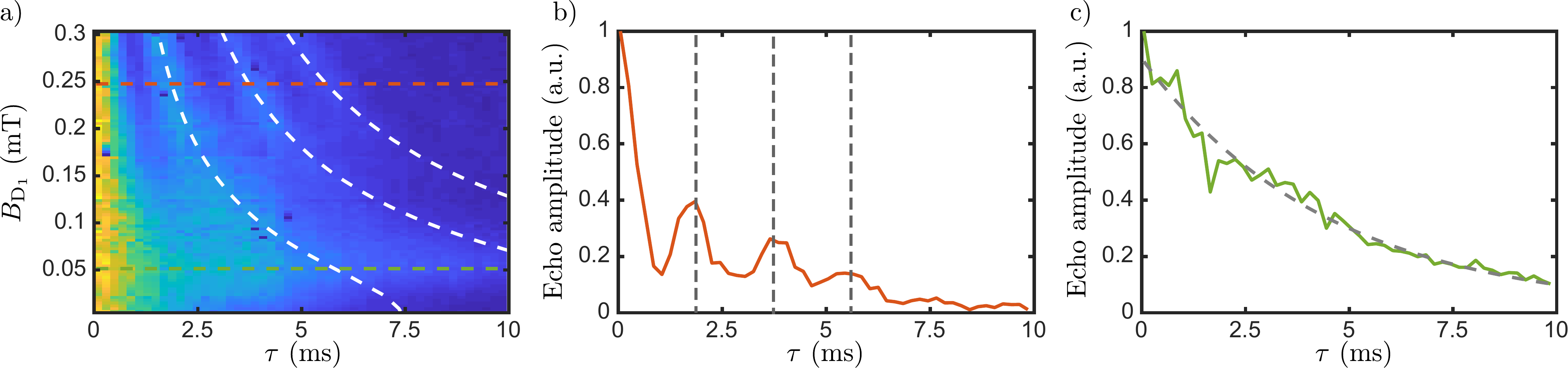}
	\caption{Map of the echo decays as a function of time delay $\tau$, for a range of fields along $B_{\text{D}_1}$ and for fixed $B_{\text{D}_2} = -220~\mu$T. The white dashed lines show multiples of the yttrium Larmor period, $n T_\text{Y}$, where $n=1,2,3$. b) The echo decay obtained at $B_{\text{D}_1} = -248~\mu$T shows clear ESEEM peaks at the multiples of $T_\text{Y}$ (vertical dashed line). c) The echo decay at $B_{\text{D}_1} = -51~\mu$T shows no discernible modulation. The magnetic fields for the two decays in b) and c) are indicated as horizontal dashed lines in a), and with circles in Fig. \ref{fig:scanD1D2}(c).}
	\label{fig:scanD1_ESEEM}
\end{figure*}

A map with finer resolution in the field was also recorded around the point of highest echo amplitude, for $\tau=5$~ms, see Fig. \ref{fig:scanD1D2}(c). A symmetric pattern appears around a field offset of around $B_{\text{D}_2} = -155~\mu$T along the D$_2$ axis. Such offsets have been observed in all our past echo measurements in \ybiso{} \cite{Ortu2018,Businger2020}, and it varies depending on the cryostat and electromagnetic environment, showing that it is produced externally from the crystal. In general, these slowly varying background fields can make it challenging to experimentally find and maintain the ZEFOZ point. In the following, we assume that the true zero field, i.e. the ZEFOZ point, is located close to $B_{\text{D}_1} = 0$ and $B_{\text{D}_2} = -155~\mu$T.

The strong ESEEM pattern seen in Fig. \ref{fig:scanD1D2}(c) indicates that the superhyperfine coupling contributes significantly to the effective spin dephasing mechanism away from the ZEFOZ point. To investigate this further, we recorded Hahn echo decay curves as a function of $\tau$, for a varying field $B_{\text{D}_1}$ along the D$_1$ axis. A constant field of $B_{\text{D}_2} = -155~\mu$T was applied in the D$_2$ axis, to compensate for the observed lab bias field. The data is presented as a 2D map in Fig. \ref{fig:scanD1D2}(d), and a few examples of decay curves are shown in Fig. \ref{fig:scanD1D2}(e). For the highest fields along D$_1$, there is a distinct ESEEM modulation appearing, with an almost complete collapse of the echo signal before the first ESEEM revival. This shows that the superhyperfine coupling plays a major role in the observed spin dephasing, as was also recently observed with photon echoes in Er-doped \yso{} \cite{Car2020}. Our measurements also suggests that coupling to $^{89}$Y$^{3+}$ ions is dominating, as the time of the first ESEEM revival closely follows the Lamor period of the yttrium spin $T_\text{Y} = 1/(B \gamma_\text{Y})$, where $\gamma_\text{Y} = 2.095$~MHz/T, see Fig. \ref{fig:scanD1D2}(d). Note that in the model, the field amplitude $B$ is taken to be the field applied along D$_1$, i.e. $B = B_{\text{D}_1}$, as the D$_2$ field only compensates the lab bias field.

The ESEEM oscillation seen in Figs \ref{fig:scanD1D2}(d)-(e) increases in period and decreases in amplitude, as the $B$ field approaches the ZEFOZ condition at the estimated zero-field point. The increased period is clearly due to the yttrium Larmor frequency going to zero, while we attribute the decrease in amplitude to the quenching of the effective \ybi{} dipole moment in the ZEFOZ point. Note that this is a very effective method for finding the true $B=0$ point experimentally, by pushing the oscillation to longer periods until its amplitude vanishes. However, as both the Larmor frequency and the first-order effective dipole moment goes to zero, it is not possible to discriminate the two effects while approaching $B=0$.

A clearer way of showing how the ESEEM oscillation can be quenched by minimizing the effective dipole moment is to approach the line of minimum $S_1$ gradient, while applying a constant offset field along D$_2$ such that the yttrium Larmor frequency is strictly non-zero for all measurements. In Fig. \ref{fig:scanD1_ESEEM}(a), we show a 2D map of Hahn echo decays, as a function of fields applied along the D$_1$ axis. The D$_2$ field was held constant at $B_{\text{D}_2} = -220~\mu$T, meaning there was a constant offset of $-65~\mu$T along D$_2$ from the estimated true $B=0$ point. The data shows up to three clear ESEEM oscillations for the highest fields along D$_1$, see the example in Fig. \ref{fig:scanD1_ESEEM}(b). However, as the D$_1$ field is tuned towards the line of minimum dipole moment, the oscillations completely vanish, although the Larmor frequency is non-zero at all times. For the field of $B_{\text{D}_1} = -51~\mu$T, which is exactly on the line of minimum gradient (cf. Fig. \ref{fig:scanD1D2}(c)), the echo amplitude displays an exponential decay without oscillations, see Fig. \ref{fig:scanD1_ESEEM}(c). The dependence of the period of oscillation on the field magnitude $B$ follows closely the Larmor frequency, shown by the theoretical lines based on the model field amplitude $B = \sqrt{ (65~\mu \text{T})^2 + B_{\text{D}_1}^2}$, see Fig. \ref{fig:scanD1_ESEEM}(a). The quenching of the ESEEM modulation at a non-zero field magnitude unambiguously demonstrates that the effective dipole moment is field dependent in \ybi{} and that it can be minimized at specific orientations. This could provide an interesting tool for studying superhyperfine interactions in detail.

\begin{figure*}[t!]
	\centering
	\includegraphics[width=0.9\textwidth]{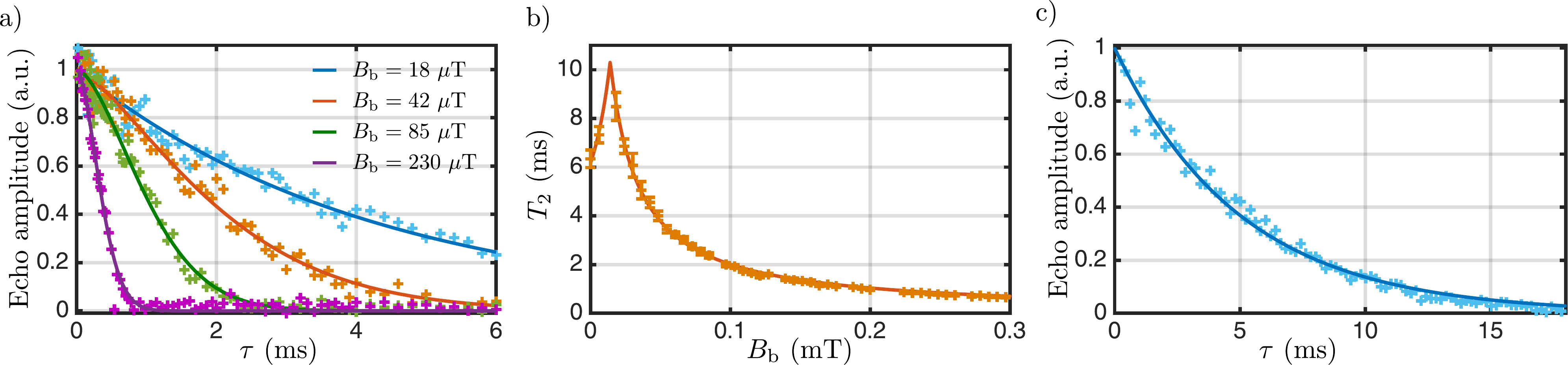}
	\caption{a) Hahn echo decays for fields $B_b = 18, 42, 85, 230~\mu$T along the $b$-axis, with a constant offset field $B_{\text{D}_2} = -155~\mu \text{T}$ to compensate for the lab bias field along D$_2$. Solid lines show fits to a stretched exponential. b) Spin coherence times measured along the $b$ axis. The solid line shows a simple empirical model (see text for details). c) The longest Hahn echo decay was obtained by carefully adjusting the bias magnetic field around the estimated ZEFOZ point, while optimizing the echo amplitude at a long delay, resulting in a spin coherence time of $T_2 = (10.0 \pm 0.4)~\text{ms}$.}
	\label{fig:scanb}
\end{figure*}

The Hahn echo decay was also measured while applying a varying field in the $b$ axis, starting from the estimated ZEFOZ point (i.e. with $B_{\text{D}_2} = -155~\mu \text{T}$ constant). Along this axis, no oscillations appear for up to $B_b = 300~\mu \text{T}$, as shown in Fig. \ref{fig:scanb}(a). However, the timescale of the decay is similar as measured along the D$_1$ axis, indicating that the effective spin dephasing time is also governed by superhyperfine coupling along the $b$-axis. The spin decay curves were fitted to a stretched exponential $E(\tau) = E(0) \exp(-(2\tau/T_2)^m)$, where $m$ is the Mims factor, see Fig. \ref{fig:scanb}(a). 

The spin coherence time decreases as the $B_b$ is increased, as shown in Fig. \ref{fig:scanb}(b), following a $1/B_b$ dependence at higher fields, as already observed in Ref. \cite{Ortu2018}. The coherence time data fits well to a simple empirical formula $T_2(B) = 1/(1/T_2(0)+\pi \kappa \abs{B-B_0})$, where $T_2(0)$ is the estimated zero-field coherence time, $\kappa$ the field dependent homogeneous linewidth, and $B_0$ the offset bias field along the b-axis. The fit to data yields $T_2(0) = (10.3 \pm 0.2)~\text{ms}$, $\kappa = (1.48 \pm 0.04) \text{MHz/T}$ and $B_0 = (14.1 \pm 0.4)~\mu \text{T}$. Similar values were found when fitting only the first collapses before the first ESEEM revival of the D$_1$ scan shown in Fig. \ref{fig:scanD1D2}(d) and (e)  (see Supplementary Information). Two aspects stand out in this simple analysis, which is the universal $1/B$ dependence (also observed in Ref. \cite{Ortu2018}) and a field-dependent homogeneous linewidth $\kappa$ that is rather close to the yttrium gyromagnetic ratio $\gamma_\text{Y}$. As it stands, we do not have a detailed model to explain these observations, but we observe that the screened superhyperfine decay model in Ref. \cite{Car2020} cannot be applied here since the dipole moment is proportional to $B$ (see Supplementary Information). In any case, it appears clearly that a model of the spin dephasing times close to zero field must include superhyperfine interactions with yttrium ions.

Finally, we briefly discuss the longest spin coherence times obtained in this study. 
As observed in our dataset,
magnetic bias fields of the order of a few $\mu \text{T}$ affect the Hahn echo decay. Exponential functions were fitted to the longest decays obtained for the D$_1$ and $b$ axis scans, see Figs \ref{fig:scanD1D2}(e) and Fig. \ref{fig:scanb}(a), from which we obtained $T_2 = (9.6 \pm 0.8)~\text{ms}$ and $T_2 = (8.5 \pm 0.6)~\text{ms}$, respectively. In a separate measurement, where we finely adjusted the bias field around the expected ZEFOZ point, we obtained $T_2 = (10.0 \pm 0.4)~\text{ms}$ as shown in Fig. \ref{fig:scanb}(c). This can be compared to the $T_2 = (6\pm 1)~\text{ms}$ obtained in a parallel study by Alexander \textit{et al.} \cite{Alexander2022} 
in a \yso{} crystal doped with 5 ppm of \ybi{}. In this context, it is also worth noting that we measured the optical coherence time in both crystals, and found a similar relative difference of $(0.610\pm 0.050)~\text{ms}$ and $(1.05\pm 0.130)~\text{ms}$ for the 5 and 2 ppm doping, respectively (see Supplementary Information).

\section{Conclusions and Discussion}

The lumped-element resonator provides a method for coherent driving of both optical and MW transitions over a large sample volume, giving us access to all the higher MW spin transitions for quantum memory schemes \cite{Businger2020}. The polarization-dependent optical transition strengths between the hyperfine levels opens up many possible $\Lambda$-schemes, particularly involving the high-frequency spin transitions such as the 1.841 GHz and 3.025 GHz transitions (see Supplementary Information), allowing one to optimize the memory scheme in terms of other parameters, such as the maximum memory bandwidth. Integrating the lumped-element resonator in a spin-wave quantum memory experiment, cf. Ref. \cite{Businger2020}, is an important next step.

The experiments presented in this work show that superhyperfine-induced ESEEM oscillations of the Hahn echo signal have a strong influence on the dephasing dynamics at low fields. We have also shown that the effective magnetic dipole moment is field-orientation dependent at low fields and can be quenched at the zero-field ZEFOZ point, or when the $S_1$ gradient is very low. These experimental observations are also well-supported by our calculations. There remains many open and interesting questions, however, such as the influence of a stochastic time-dependent spectral diffusion process \cite{Lange2010}, which has been successfully applied to europium-doped \yso{} \cite{Holzaepfel2020,Ortu2022}. The simple empirical model to explain the $B$-field dependence of the coherence time includes a zero-field $T_2(0)$, which could be a sign that the observed zero-field coherence time could be dominated by spectral diffusion. In addition, one would expect to see more distinct ESEEM oscillations at longer delays, as seen in Ref. \cite{car2020superhyperfine}, which would also suggest another mechanism at long time scales. Since the effective magnetic dipole moment can be tuned by the strength and orientation of the magnetic field, we believe that \ybiso{} provides a particularly interesting system for further studies of these dephasing mechanisms.

\section{Methods}

The magnetic fields along the D$_1$ and D$_2$ axes were generated by two pairs of coils, placed outside the cryostat. A superconducting coil placed inside of the cryostat generated the field along the $b$-axis. The magnetic fields were calibrated by measuring the Zeeman frequency shifts they induced, and comparing these to the shifts calculated with the model presented in \cite{Tiranov2018}.

To increase the signal of the RHS echo, the spin echo sequence is preceded by a population preparation step, in order to increase the population difference between states $\ket{4_g}$ and $\ket{2_g}$. To do so, the laser is first tuned to the $\ket{2_g}$-$\ket{1_e}$ transition to decrease the population in the $\ket{2_g}$ state. Then, the frequency of the laser is scanned to address optical transitions involving the $\ket{1_g}$, $\ket{2_g}$ and $\ket{3_g}$ states, thereby pumping ions into the state $\ket{4_g}$.

To drive the $\ket{2_g}$-$\ket{4_g}$ MW-transition with high power, a standard wifi amplifier was used (Sunhans SH24Gi20W).

\section*{DATA AVAILABILITY}

The datasets shown in the figures of this article can be accessed from the Zenodo data repository 10.5281/zenodo.7064041.

\section*{ACKNOWLEDGEMENTS}

We acknowledge funding from the Swiss National Science Foundation (SNSF) through project 197168, the French Agence Nationale de la Recherche through project MIRESPIN (Grant No. ANR-19-CE47- 0011) and support from DGA. 

\section*{AUTHOR CONTRIBUTIONS}

A.T. and T.S.M. designed and made initial tests of the loop-gap resonator. The crystal growth and initial optical spectroscopy was done by E.L.H., A.F. and P.G.. All the optically detected spin resonance measurements were carried out by L.N., with support from M.B. and T.S.M.. L.N. did all the data reduction and analysis, with support from M.A. The theoretical superhyperfin calculations were done by M.A. and T.C.. The manuscript was mainly written by N.L. and M.A., with contributions from all the authors. M.A. provided overall oversight of the project.

\section*{COMPETING INTERESTS}

The authors declare no competing interests.

\bibliography{mw_ybyso.bib}

\end{document}